\documentclass[a4paper,12pt]{article}

\usepackage{amsmath}
\usepackage{amssymb}
\usepackage[dvips]{graphicx}
\usepackage[dvips]{psfrag}

\makeatletter
\@addtoreset{equation}{section}
\renewcommand{\theequation}{\thesection.\@arabic\c@equation}
\makeatother

\makeatletter
\renewcommand\appendix{\par
  \setcounter{section}{0}%
  \setcounter{subsection}{0}%
  \gdef\thesection{Appendix \@Alph\c@section }
  \renewcommand{\theequation}
  {\Alph{section}.\arabic{equation}}
}
\makeatother

\begin{document}

\begin{titlepage}

\vspace*{-15mm}   
\baselineskip 10pt   
\begin{flushright}   
\begin{tabular}{r}    
{\tt KUNS-2434}\\ 
{\tt KEK-TH-1605}\\
February 19, 2013
\end{tabular}   
\end{flushright}   
\baselineskip 24pt   
\vglue 10mm   

\begin{center}
{\Large\bf
 A Self-consistent Model of \\
 the Black Hole Evaporation
}

\vspace{8mm}   

\baselineskip 18pt   

\renewcommand{\thefootnote}{\fnsymbol{footnote}}

Hikaru~Kawai$^a$\footnote[2]{hkawai@gauge.scphys.kyoto-u.ac.jp}, 
Yoshinori~Matsuo$^b$\footnote[3]{ymatsuo@post.kek.jp} and 
Yuki~Yokokura$^a$\footnote[4]{yokokura@gauge.scphys.kyoto-u.ac.jp}

\renewcommand{\thefootnote}{\arabic{footnote}}

\vspace{5mm}   

{\it  
 $^a$ Department of Physics, Kyoto University, 
 Kitashirakawa, Kyoto 606-8502, Japan \\
 $^b$ KEK Theory Center, 
 High Energy Accelerator Research Organization(KEK), \\
 Oho 1-1, Tsukuba, Ibaraki 305-0801, Japan
}
  
\vspace{10mm}   

\end{center}

\begin{abstract}
We construct a self-consistent model which describes a black hole from formation to evaporation 
including the back reaction from the Hawking radiation. 
In the case where a null shell collapses, 
at the beginning the evaporation occurs, 
but it stops eventually, and a horizon and singularity appear. 
On the other hand, in the generic collapse process of a continuously distributed null matter, 
the black hole evaporates completely  without forming a macroscopically large horizon nor singularity. 
We also find a stationary solution in the heat bath, 
which can be regarded as a normal thermodynamic object. 
\end{abstract}

\baselineskip 18pt   

\end{titlepage}

\newpage
\section{Introduction}\label{sec:Intro}
In the analysis of the black hole evaporation,
one usually assumes that a horizon is formed in a collapse process, 
and examines the evaporation and entropy in the static black hole \cite{Bekenstein:1973ur}-\cite{Bianchi:2012br}. 

In this paper we try to build a self-consistent model which describes both formation and evaporation of a black hole 
including the back reaction from the Hawking radiation\footnote{Note that 
we mean by ``black hole" not one that has an event horizon defined globally as in the rigorous sense, 
but one that is formed in a semi-classical collapse process.
Some authors pursued similar ideas \cite{Boulware}-\cite{Stephens:1993an}.
}. 
That is, we solve the semi-classical Einstein equation in a self-consistent manner:
\begin{equation}
G_{\mu \nu}=8\pi G \langle T_{\mu \nu} \rangle,
\label{s_E}
\end{equation}
where $\langle T_{\mu \nu} \rangle$ contains the contribution from both the collapsing matter and the Hawking radiation. 
From the solution we can investigate whether a horizon and singularity are formed or not.

We first consider a null shell as the collapsing matter and construct 
the geometry by connecting the inside flat metric and the outside outgoing Vaidya metric on the shell.
Note that particle creation generally occurs in a time-dependent gravitational potential, 
and especially, the Hawking radiation can appear without a horizon \cite{Barcelo:2010xk}. 
We invent a formula that evaluates the energy flux of such a process. 
Then we obtain self-consistent equations which determine time evolution of the shell and the radiation.
The solution shows that 
the radiation stops, the horizon and singularity appear, and the black hole remains forever.

Next we analyze the case where a continuous null matter collapses and 
discuss the mechanism of the Hawking radiation. 
It has an onion-like internal structure and evaporates gradually from the outermost part.
Then we write down a self-consistent stationary solution in the heat bath.
It has neither a macroscopically large horizon nor singularity.

\section{Construction of a model}\label{sec:model}
We first explain the general idea for construction of a geometry which describes a black hole from formation to evaporation.
Next we propose a simple model.

\begin{figure}
 \begin{center}
 \includegraphics*[scale=0.22]{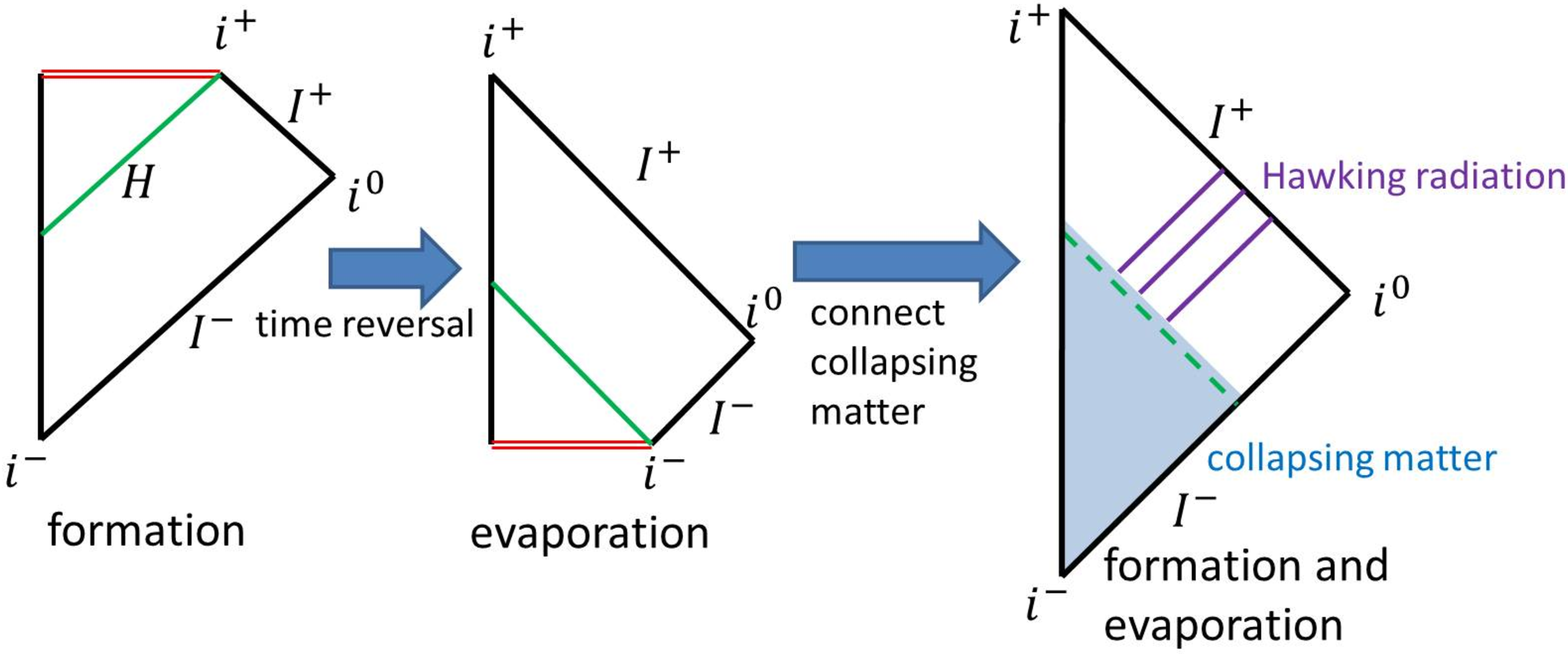}
 \caption{Penrose diagrams corresponding to  the general idea for construction of a geometry which describes 
 both formation and evaporation of a black hole.
The left one represents the formation of a Schwarzschild black hole which has the event horizon and singularity. 
The center one corresponds to the evaporation from the Schwarzschild black hole to the flat spacetime.
The right one describes both the formation and evaporation.}
 \label{fig:general}
 \end{center}
 \end{figure}

Suppose that a gravitational collapse forms a Schwarzschild black hole as in the left of Fig.\ref{fig:general}. 
If we take time reversal, the existing black hole goes back to the flat spacetime as in the center of Fig.\ref{fig:general}.
Then, if we cover the inside of the horizon and the singularity 
by pasting a collapsing matter, 
we obtain a geometry which describes both the formation and evaporation as in the right of Fig.\ref{fig:general}
\footnote{In \cite{Stephens:1993an} a similar diagram is discussed.}. 
Note that whether this picture is realized or not depends on the dynamics.
Therefore we need to make some model and solve it concretely.

We will consider the following model.
When we take a null shell as the collapsing matter, the inside spacetime is flat:
\begin{equation}
 ds^2 = -d U^2 - 2 d U dr + r^2 d\Omega^2 .
 \label{flat}
\end{equation}
As a simple model of the outside metric, we take the outgoing Vaidya metric \cite{Vaidya}:
\begin{equation}
 ds^2 = -\left(1-\frac{a(u)}{r}\right) du^2 - 2 du dr + r^2 d\Omega^2 , 
 \label{Vaidya}
\end{equation}
where $m(u)=\frac{a(u)}{2G}$ is the Bondi mass and 
the only non-zero component of the Einstein tensor is 
\begin{equation}
G_{uu}=-\frac{\dot{a}(u)}{r^2},
\label{G_uu}
\end{equation}
where the null energy condition implies $\dot a <0$
\footnote{In \cite{Hiscock:1980ze} and \cite{Girotto:2004iu}, the ingoing Vaidya metric was used to study the evaporation.}.
This is the general spherically symmetric metric which satisfies $G^{\mu}{}_{\mu}=0$ and $G_{\mu\nu}=0$ except for $G_{uu}$ 
\footnote{These conditions come from the following discussion.
At $r\gg a$, where $a$ is the Schwarzschild radius of the null shell, we can take $G_{\theta \theta}=G_{\phi \phi}=0$ 
because most partial waves with $l\gg 1$ of the radiation do not go through their own centrifugal barrier in $V_l \sim \frac{l(l+1)}{r^2}$.
Next the incoming flux can be neglected there 
because of the boundary condition that any energy flow does not come from infinity except for the shell.
Furthermore if we consider only massless fields, we can assume $G^{\mu}{}_{\mu}=0$ 
because the Weyl anomaly vanishes approximately in $r\gg a\gg l_p$.
At $r \sim a$, the ingoing flow and $T_{\theta \theta}$ can exist with $l\gg 1$, 
but we assume to neglect them for the simplest model.
Therefore we can consider the conditions.
In this sense the outgoing Vaidya metric represents the outgoing radiation without the gray-body factor.}.

\begin{figure}
 \begin{center}
 \includegraphics*[scale=0.135]{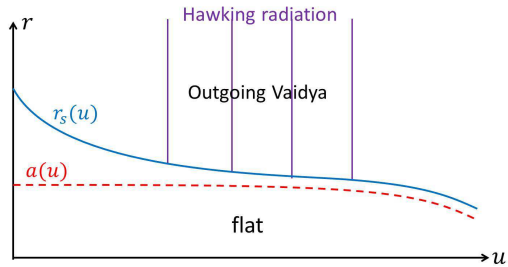}
 \caption{One-shell model. The outgoing Vaidya metric and the flat metric are connected by the null shell. 
 The null shell $r_s(u)$ approaches its own Schwarzschild radius $a(u)$ exponentially.}
 \label{fig:Vaidya}
 \end{center}
 \end{figure}

Note that the coordinate $r$ must be the same in the both side, 
because it is defined as the radius of 2-sphere and there is no room to rewrite $r^2 d\Omega^2$.
On the other hand, $u$ is related to $U$ as 
\begin{equation}
dU = -2dr_{s} =\left(1-\frac{a(u)}{r_{s}(u)}\right) du , 
\label{junction}
\end{equation}
where $r_s (u)$ is the locus of the null shell. 
This comes from the fact that $r_s$ is an ingoing null geodesic in the both sides. 
Thus a simple model is given by connecting the outgoing Vaidya metric and the flat metric with the null shell as in Fig \ref{fig:Vaidya}. 
We call it one-shell model.

Here we analyze the locus of the null shell $r_s(u)$ for a given function $a(u)$.
$r_s(u)$ is determined by the condition \eqref{junction}: 
\begin{equation}
\frac{dr_s(u)}{du}=-\frac{r_s(u)-a(u)}{2r_s(u)}.
\label{r_eq}
\end{equation}
This equation tells that the shell will approach its own Schwarzschild radius in the time scale $\sim a$ 
if $a(u)$ changes so slowly that the time scale in which $a(u)$ changes significantly, $\frac{a}{|\dot a|}$, is much larger than $a$, that is,
\begin{equation}
|\dot a|\ll 1
\label{a_slow}
\end{equation}
Then, in the region $r_s \sim a$, we can replace $r_s$ in the denominator with $a$ and solve it as 
\begin{equation}
r_s(u)\approx a(u)-2a(u)\dot{a}(u)+Ca(u)e^{-\frac{u}{2a(u)}},
\label{r_s}
\end{equation}
where $C$ is a positive constant. 
Here the term $-2a\dot{a}$ means that 
as the shell approaches to its Schwarzschild radius in the time scale of $2a$, 
the radius reduces by the evaporation. (See Fig~\ref{fig:Vaidya}.) 
Therefore the shell cannot catch up with the radius completely as long as $\dot a <0$, 
but it approaches to 
\begin{equation}
r_s(u)\approx a(u)-2a(u)\dot{a}(u). 
\label{r_asy}
\end{equation}

Finally we investigate the surface energy-momentum tensor $T_{\Sigma}^{\mu\nu}$ on the shell. 
Using the Barrabes-Israel null-shell formalism \cite{Barrabes:1991ng, Poisson}, \ we estimate 
\begin{align}
 T^{\mu\nu}_{\Sigma} &= (-k\cdot v)^{-1}\delta (\tau) \left(  \mathcal M k^{\mu} k^{\nu} + \mathcal P \sigma^{\mu\nu} \right) ,\\
 \mathcal M &= \frac{a}{8\pi G r_s^2},\qquad 
 \mathcal P = \frac{-\dot{a}r_s}{4\pi G(r_s-a)^2}, 
 \label{MassOnLocus}
\end{align}
where 
$v=\frac{\partial}{\partial \tau}$ is the four vector of an observer ($v^2=-1$), 
$k^{\mu}$ is the ingoing radial null vector which is taken as $k^{\mu}\partial_{\mu}=\frac{2}{1-\frac{a(u)}{r}}\partial_u-\partial_r$ in the Vaidya metric \eqref{Vaidya} and 
$k^{\mu}\partial_{\mu}=2\partial_{U}-\partial_r$ in the flat space \eqref{flat}, 
and $\sigma^{\mu\nu}$ is the metric on the 2-sphere ($\sigma_{\mu\nu}dx^{\mu}dx^{\nu}=r^2d\Omega^2$). 
The fact that $\mathcal P \propto -\dot a(u)>0$ implies that 
the work done by the shell as it contracts is transformed to the Hawking radiation. 
Thus this model is consistent in energetics.

Time evolution of this model depends on the functions $a(u)$ and $r_s(u)$, 
so we will investigate their dynamics in the following sections.

\section{Flux formula}\label{sec:flux}
We will here construct a flux formula $J(u)$ which, at $r\gg a$, estimates energy flow from the black hole:
\begin{equation}
\frac{dm}{du}=-J(u).
\label{balance}
\end{equation}
In the Heisenberg picture, we use the Eikonal approximation, the point-splitting regularization and only the s-wave to obtain
\begin{equation}
J(u)=\frac{\hbar}{8\pi}\left[ \frac{\ddot U(u)^2}{\dot U(u)^2}-\frac{2\dddot U(u)}{3\dot U(u)}\right]\equiv \frac{\hbar}{8\pi}\left\{u,U\right\},
\label{J}
\end{equation}
whose form is the same as the Schwarzian derivative.
The derivation is given in the \ref{sec:der_flux}.
Note that we can also derive the Planck distribution without horizon (see \ref{sec:Planck}) .

First we test the formula in the case without back reaction, that is, 
in the geometry obtained by connecting the Schwarzschild metric and the flat space.
In this case, from \eqref{r_s}, $r_s(u)$ becomes 
\begin{equation}
r_s(u) = a+Cae^{-\frac{u}{2a}},
\label{r_s2}
\end{equation}
where $a$ becomes constant completely. 
Then the flux is estimated as 
\begin{equation}
J(u)=\frac{\hbar}{96\pi}\frac{1}{a^2}=\frac{\pi}{6\hbar}T_H^2,
\label{J_const}
\end{equation}
where $T_H=\frac{\hbar}{4\pi a}$.
This is the same as thermal radiation from a one-dimensional black body with the temperature $T_H$.
In this sense, the flux formula \eqref{J} is consistent with the usual result 
\cite{Hawking:1974sw}. 
Note that this result is the same as the Stefan-Boltzmann law except for the coefficient. 

From \eqref{junction}, \eqref{r_eq}, \eqref{balance} and \eqref{J}, 
we have obtained the self-consistent equations which determine the dynamics of the one-layer model, 
that is, $a(u)$ and $r_s(u)$: 
\begin{equation}
\frac{dr_s}{du}=-\frac{r_s(u)-a(u)}{2r_s(u)},
\label{sceq1}
\end{equation}
\begin{equation}
\frac{da}{du}=-2GJ(u)=-\frac{l_p^2}{4\pi}\left[ \frac{\ddot r_s(u)^2}{\dot r_s(u)^2}-\frac{2\dddot r_s(u)}{3\dot r_s(u)}\right],
\label{sceq2}
\end{equation}
where $l_p=\sqrt{G\hbar}$ is the Planck length.

\section{Time evolution of a null shell}\label{sec:thin}
\begin{figure}
\begin{center}
\includegraphics*[scale=0.19]{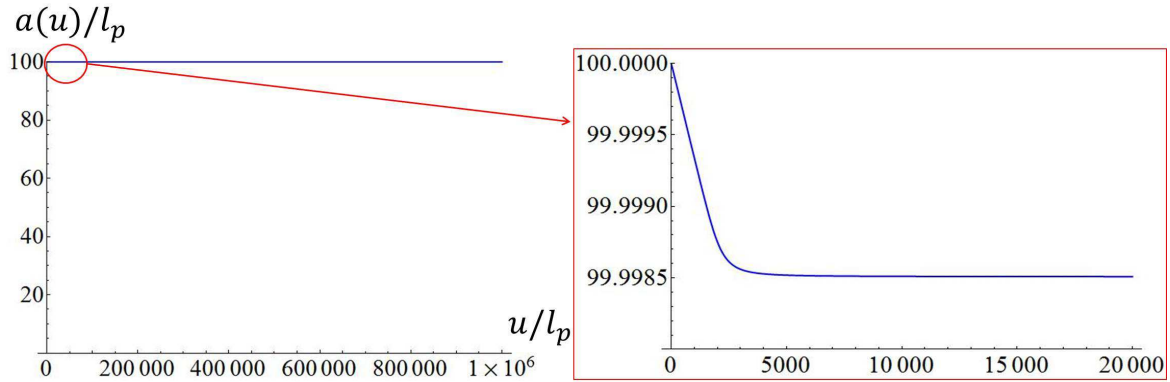}
\caption{The numerical result of $a(u)$ in the case of the null shell.
The initial conditions are given by \eqref{initial} with $a(0)=100$ and $n=10/9$.}
\label{fig:shell}
\end{center}
\end{figure}

Now we consider the collapse of a null shell by using the one-shell model 
and investigate whether it evaporates or not 
\footnote{A similar case was studied in a different set up \cite{Callan:1992rs} \cite{Susskind:1993if}.}.
The numerical result of \eqref{sceq1} and \eqref{sceq2} is shown in Fig.~\ref{fig:shell}. 
Here we have chosen the initial conditions given by 
\begin{equation}
r_s(0)=na(0),\quad \dot r_s(0)=-\frac{n-1}{2n}, \quad \ddot r_s(0)=0,
\label{initial}
\end{equation}
where $n\gtrsim 1$ is a number, and we assume $a(0)\gg l_p$. 
The shell does not evaporate completely, and a horizon and singularity appear. 
This asymptotic behavior does not depend on the detail of the initial conditions. 

We can understand why the radiation stops in the following manner. 
Let's recall the estimation of the Hawking radiation on the geometry with $\dot a =0$ (see \eqref{r_s2}). 
In that case, only the term proportional to $e^{-\frac{u}{2a}}$ contributes to the formula. 
However, now the term $\dot a$ appears in $r_s$, \eqref{r_s}, 
and the exponential factor will damp for large $u$. 
Therefore, $r_s$ becomes $a-2\dot a a$ asymptotically as in \eqref{r_asy}. 
Because $\dot a a $ is at most of order $l_p^2/a \ll l_p$, we can approximate 
\begin{equation}
r_s(u) \approx a(u).
\label{ra}
\end{equation}
Then \eqref{sceq1} and \eqref{sceq2} can be solved as 
\begin{eqnarray}
u &=& \frac{e^{-\frac{D^2}{2}}}{6\pi B}\int^{\xi}_D d\xi'e^{\frac{1}{4}\xi'{}^2}~,\label{u_eq}\\
a(u) &=& a(0) -B \int^{\xi}_D d\xi'e^{-\frac{1}{4}\xi'{}^2}~,
\label{a_eq}
\end{eqnarray}
where $B$ and $D$ are integration constants, and $B$ is small and positive. 
From \eqref{u_eq} $u=\infty$ corresponds to $\xi = \infty$, 
so \eqref{a_eq} shows that $a(u)$ will not necessarily vanish as $u\rightarrow \infty$. 

Thus we have seen that a collapsing null shell with radius $r_s$ radiates for a while, 
but it stops and the radius $r_s$ almost stays at the Schwarzschild radius $a$. 
Then the horizon and singularity appear. 
A single shell does not evaporate completely 
even if the  back reaction from the Hawking radiation is taken into account.

\section{Generalization to a continuous null matter and the stationary solution}\label{sec:general}
\begin{figure}
\begin{center}
\includegraphics*[scale=0.18]{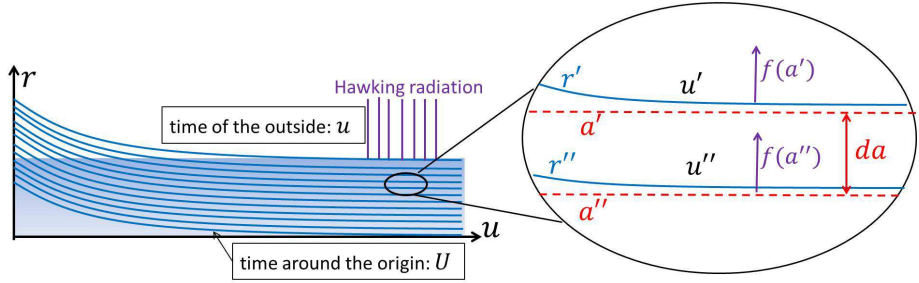}
\caption{Collapse of a continuous null matter to the asymptotic region 
where \eqref{r_asy} holds for each shell. 
The vicinity around a point behaves as in the one-shell model in its own time coordinate $u'$. 
The redshift factor is required to compare physics between different shells.
}
\label{fig:cont}
\end{center}
\end{figure}

We discuss the case where a continuous null matter collapses (see Fig.~\ref{fig:cont}). 

Let's consider a shell. 
The metric just outside the shell is given by 
\begin{equation}
 ds^2 = -\left(1-\frac{a'(u')}{r}\right) du'^2 - 2 du' dr + r^2 d\Omega^2 , 
 \label{Vaidya'}
\end{equation}
where $a'(u')$ is the Schwarzschild radius corresponding to the total energy of the lower shells, 
and $u'$ is the time coordinate for the shell. 
The locus of the shell $r'(u')$ follows 
\begin{equation}
\frac{dr'(u')}{du'}=-\frac{r'(u')-a'(u')}{2r'(u')}~,
\label{r_eq'}
\end{equation}
and behaves as in the one-shell model:
\begin{equation}
r'(u')\approx a'-2a'\frac{da'}{du'}+C'a'e^{-\frac{u'}{2a'}}~.
\label{r'}
\end{equation}
Furthermore, for the shell, the flux formula holds, and we have 
\begin{equation}
\frac{da'}{du'}=-2GJ'=-\frac{Nl_p^2}{4\pi}\left\{u', U \right\}~,
\label{J'}
\end{equation}
where $J'$ represents the energy flux measured at infinity if the shells outside did not exist.
Here we have introduced $N$ degrees of freedom.
In the case of the standard model, $N\sim 100$ in the energy region higher than the weak scale. 
Note that introducing $N$ corresponds to replacing $l_p$ with $\sqrt{N}l_p$.
\subsection{Hawking radiation from each shell}
We will show that the Hawking radiation is emitted from each shell, 
but only shells near the outermost one are relevant because of the large redshift.

First we introduce a notation 
\begin{equation}
\rho' \equiv r'-a'~,
\label{rho}
\end{equation}
which represents distance between the shell $r'$ and the Schwarzschild radius $a'$. 
Here we assume that the last term in \eqref{r'} has already damped for each shell, 
so that we have 
\begin{equation}
\rho' = -2a'\frac{da'}{du'}~.
\label{rho2}
\end{equation}
Then the energy flux for each shell depends only on its Schwarzschild radius 
and does not have an explicit $u'$-dependence: 
\begin{equation}
\frac{da'}{du'}=-f(a')~.
\label{assume}
\end{equation}
From \eqref{rho2}, $\rho '$ also becomes a function of $a'$:
\begin{equation}
\rho(a') = 2a' f(a')~.
\label{rho3}
\end{equation}

Now we consider the junction condition of the adjacent shells. 
By looking at each shell from the both side (see Fig.~\ref{fig:cont}), we obtain 
\begin{equation}
\frac{r'-a'}{r'}du'=\frac{r'-a''}{r'}du''.
\label{junction'}
\end{equation}
If $a'-a''=da$ is small, we get
\begin{equation}
\frac{du'}{du''}=\frac{r'-a''}{r'-a'}=\frac{\rho'+da'}{\rho'}=1+\frac{da'}{\rho'}.
\end{equation}
By integrating it, we obtain the redshift factor between 
$a'$ and $a''$ for finite distance: 
\begin{equation}
\frac{du'}{du''}=\exp \left(\int^{a'}_{a''}\frac{d\bar a}{\rho (\bar a)}\right).
\label{red}
\end{equation}
From this equation, the quantity 
\begin{equation}
\xi' \equiv \frac{d}{du'} \log \left( \frac{dU}{du'}\right)
\label{xidef}
\end{equation}
can be rewritten as 
\begin{equation}
\xi' = \frac{d}{du'}\left( - \int^{a'(u')}_{0}\frac{d\bar a}{\rho (\bar a)} \right)
=-\frac{da'}{du'}\frac{1}{\rho(a')}=\frac{1}{2a'},
\end{equation}
where  \eqref{rho2} and \eqref{rho3} have been used. 
Thus we have found that for each shell 
\begin{equation}
\xi'=\frac{1}{2a'},
\label{xi}
\end{equation}
which is independent of a concrete form of $f(a')$.

On the other hand, by expressing $\{u',U\}$ in terms of $\xi'$, 
we can express the energy flow in \eqref{J'} as 
\begin{equation}
J'=\frac{N\hbar}{24\pi}\left( \xi^{'2}-2\frac{d\xi'}{du'} \right)~. 
\label{J'2}
\end{equation}
From \eqref{xi} and \eqref{J'2}, we obtain
\begin{equation}
J'=\frac{N\hbar}{96\pi}\left(\frac{1}{a'{}^2}+\frac{4}{a'{}^2}\frac{da'}{du'} \right)~.
\label{J'4}
\end{equation}
By substituting \eqref{J'} and \eqref{J'4} iteratively, we have 
\begin{equation}
J'(u')=\frac{N\hbar}{96\pi a^{'2}} - \frac{N^2\hbar l_p^2}{1152\pi^2a^{'4}} + {\cal O}(a^{'-6}) \approx \frac{N\pi}{6\hbar}T^{'2}_H,
\label{J'3}
\end{equation}
where $T'_H=\frac{\hbar}{4\pi a'}$, 
which would be the temperature measured at infinity if the shells outside did not exist. 
Thus, any shell can emit the Hawking radiation 
if the shells are continuously distributed so that we can use \eqref{red}. 
This result does not depend on the behaviour of the shells outside the one we are considering. 
From \eqref{J'}, \eqref{rho2} and \eqref{J'3}, $\rho'$ is determined as 
\begin{equation}
\rho(a')=\frac{Nl_p^2}{24\pi a'}+{\cal O}(a^{'-3}).
\label{rho''}
\end{equation}

By considering the outermost shell, we find that the total Hawking radiation is given by 
\begin{equation}
J(u)=\frac{N\pi}{6\hbar}T^{2}_H + {\cal O}(a^{-4}),
\label{J_out}
\end{equation}
which coincides with the result for the static Schwarzschild geometry \eqref{J_const}. 
Then by applying \eqref{J'} to the outermost shell, we obtain the time evolution of the size of the black hole: 
\begin{equation}
\frac{da}{du}=-2GJ(u)=-\frac{Nl_p^2}{48\pi a^2}+{\cal O}(a^{-4})~. 
\label{a_evo}
\end{equation}
We can also show that 
the energy spectrum of the radiation follows the Planck distribution (see \ref{sec:Planck}). 
Therefore this black hole evaporates completely as is usually expected. 
However, our model describes how it happens more precisely. 
Actually the black hole evaporates gradually from the outermost shell as if one peels off an onion. 

Here we will check that the total radiation \eqref{J_out} 
is equal to the sum of the radiation from each shell. 
First let's estimate the radiation from the region between $a'$ and $a''=a'-da$ 
as depicted in Fig.~\ref{fig:cont}. 
If there were no shells outside this region, the radiation is estimated as 
\begin{eqnarray*}
f(a')-f(a'-da)e^{-\int^{a'}_{a'-da} \frac{d\bar a}{\rho (\bar a)}} &\approx& f(a') - [f(a')-\frac{d f(a') }{ da'}da]\left( 1-\frac{da}{\rho(a')}  \right)\\
 &=& \frac{d f (a')}{ da'}da + f(a')\frac{da}{\rho(a')}\\
 &\approx& \frac{da}{2a'},
\end{eqnarray*}
where $\frac{d f (a')}{ da'}da$ is neglected as a higher term, and \eqref{rho3} is used.
By using this and the redshift factor, 
the sum of radiation from each layer is estimated as 
\begin{eqnarray}
\text{the sum of radiations } &=& \int^a_0\frac{da'}{2a'}e^{-\int^a_{a'}\frac{da''}{\rho(a'')}}\nonumber \\
 &\approx& \frac{1}{2a} \int^a_0 da' e^{-\frac{a-a'}{\rho(a)}}\nonumber \\
 &=& \frac{\rho(a)}{2a} (1- e^{-\frac{a}{\rho(a)}})\nonumber \\
 &\approx& \frac{\rho(a)}{2a} \nonumber \\
 &=& f(a) = \text{the total radiation}~.
\end{eqnarray}
Here the dominant contribution in the integration comes from the outermost thin region with a width 
about $\rho (a)\propto a^{-1}$ (see \eqref{rho''}). 
Although each shell radiates, 
the outermost region gives the main contribution because of the large redshift. 
\subsection{The stationary metric}
We consider the case that the black hole is put in the heat bath 
with the Hawking temperature of the outermost shell for long time 
so that \eqref{r_asy} holds for each shell. 
It is not difficult to calculate the metric for this stationary geometry, 
and we obtain (see \ref{sec:metric}) 
\begin{equation}
ds^2=-\frac{Nl_p^2r^2}{24\pi a^4}e^{-\frac{24\pi}{Nl_p^2}(a^2-r^2)}dt^2 + \frac{24\pi r^2}{Nl_p^2}dr^2 + r^2 d\Omega^2~.
\label{metric_t}
\end{equation}
This expression is valid for $r \leq a+\frac{Nl_p^2}{24\pi a}$ 
and smoothly connected to the Schwarzschild metric at $r=a+\frac{Nl_p^2}{24\pi a}$.
This metric does not have a horizon. 
Here $t$ is the time of the flat space at infinity, which is related to the time around the origin $T$ as 
\begin{equation}
dT=\frac{N l_p^2}{a^2}e^{-\frac{12\pi}{Nl_p^2}a^2}dt~.
\label{tT}
\end{equation}
This means that $T$ is so much redshifted 
that $T$ is almost frozen from viewpoint of an observer at infinity. 
Note that this geometry has been obtained self-consistently, so the classical limit $(\hbar \rightarrow 0)$ does not exist.

This metric does not have a large curvature compared with $l_p^{-2}$ 
in the region $r\gg \sqrt N l_p$ if $N$ is sufficiently large, $N\gg 100$: 
\begin{eqnarray}
R &=& -\frac{48\pi}{Nl_p^2}-\frac{6}{r^2}\sim \frac{100}{Nl_p^2}~,\\
R_{\mu\nu}R^{\mu\nu} &=& \frac{1152\pi^2}{N^2l_p^4}+\frac{N^2l_p^4}{48\pi^2r^8}-\frac{Nl_p^2}{6\pi r^6}+\frac{16}{r^4}+\frac{288\pi}{Nl_p^2r^2}\sim \frac{10000}{N^2l_p^4}~,\nonumber \\
R_{\alpha \beta \gamma \delta} R^{\alpha \beta \gamma \delta}&=& \frac{2304\pi^2}{N^2l_p^4}+\frac{N^2l_p^4}{24\pi^2r^8}-\frac{Nl_p^2}{3\pi r^6}+\frac{20}{r^4}+\frac{384\pi}{Nl_p^2r^2}\sim \frac{10000}{N^2l_p^2}~.\nonumber
\end{eqnarray}
The singularity around the origin $r\sim 0$ is controllable in the sense that 
it can be removed by introducing a small shell surrounding the origin. 
For example, suppose a small shell with $a_0\sim C \sqrt N l_p$ comes first, 
and next, it grows to a large size with $a \gg l_p$ in the heat bath. 
Then the outside region $r>a_0$ is described as the stationary metric \eqref{metric_t}, 
while the center shell is the Schwarzschild black hole with the radius $a_0$ 
which does not evaporate forever as in the case of the one-shell model. 
Therefore we have a horizon and singularity around the origin, but their size is small. 

Here we make a comment on the Weyl anomaly.
The trace of the Einstein tensor is given by 
\begin{equation}
G^{\mu}{}_{\mu}=6\left(\frac{8\pi}{N l_p^2}+\frac{1}{r^2}\right). 
\label{G_trace}
\end{equation}
Because classically the energy-momentum tensor of null shells should be traceless, 
this should be identified with the Weyl anomaly. 
Actually, if we use the formula of the Weyl anomaly for N scalar fields \cite{Birrell}, we obtain
\begin{equation}
8\pi G\langle T^{\mu}{}_{\mu}\rangle \approx \frac{32}{15}\left(\frac{9\pi}{N l_p^2}+\frac{2}{r^2}\right), 
\end{equation}
which agrees with \eqref{G_trace} up to numerical coefficients.
Therefore the self-consistent solution obtained by the Eikonal approximation \eqref{metric_t} 
already contains the effect of the Weyl anomaly. 
\section{Conclusion and Discussion}\label{sec:CD}
We have solved the semi-classical Einstein equation in a self-consistent manner. 
We have built a model which describes a black hole from formation to evaporation 
including the back reaction from the Hawking radiation. 
We consider null matter collapse and 
assume that the geometry is obtained by connecting the matter region and the outgoing Vaidya metric. 

Using the Eikonal approximation, 
we have found a formula that gives the energy flux of the particle creation in a dynamical geometry. 
Then we have obtained the self-consistent equations which determine time evolution of the collapsing matter and radiation.

As the first example, we have analyzed the case where a single shell collapses and solved it numerically and analytically.
The shell does not evaporate completely, and a horizon and singularity appear. 
This is not a thermodynamic object but a stable one in the sense that 
it cannot be formed nor evaporated adiabatically in a heat bath. 

Next we have discussed the case where a continuous null matter collapses. 
Then the Hawking radiation occurs not only from the surface but also from the inside.
However, because of the large redshift, the radiation is emitted substantially only from the region around the surface. 
This black hole evaporates as is usually thought. 
We then have put it in a heat bath and found the stationary metric. 
It dose not have a macroscopically large horizon or singularity. 
By introducing a small shell around the origin, this singularity can be controlled. 
The metric automatically takes into account the effect of the Weyl anomaly. 

There remain some open problems.
Our stationary solution has neither horizon nor singularity,
so the information inside the hole must come back after evaporation.
However, we don't understand the mechanism clearly yet.
For example, suppose that we throw a newspaper into the stationary black hole described by our metric.
It will behave like another null shell going to the hole as it approaches the surface.
Clearly its energy will be transformed into the Hawking radiation by our mechanism.
However, the radiation itself comes from the quantum field on the past infinity, or the vacuum.
How will the information of the newspaper come back?
A clue to this problem is that we have taken the expectation value of the energy-momentum tensor $\langle T_{\mu\nu} \rangle$ 
in our self-consistent equations, 
which might correspond to the coarse-graining procedure in the ordinary statistical mechanics. 

On the other hand, if we put our black hole in a heat bath with the temperature equal to the Hawking 
temperature of the outermost shell, it is completely stationary. 
In this sense, our black hole can be regarded as a thermodynamic object having this temperature 
and its entropy is given by the area law. 
We don't claim that the information problem is solved, 
but our black hole does not have a macroscopically large horizon and singularity. 
The small singularity around the origin would be resolved by string theory. 
If it is the case, the system is completely well-defined. 


\section*{Acknowledgments}
The work of Y.M. is supported by the JSPS Research Fellowship for Young Scientists. 
The work of Y.Y. is supported by the JSPS Research Fellowship for Young Scientists 
and by the Grant-in-Aid for the Global COE Program 
``The Next Generation of Physics, Spun from Universality and Emergence'' from the 
Ministry of Education, Culture, Sports, Science and Technology (MEXT) of Japan.

\appendix 
\section{Derivation of the flux formula \eqref{J}} \label{sec:der_flux}
We will here derive the flux formula \eqref{J} by taking only the s-wave and using the Eikonal approximation.
From \eqref{s_E} and \eqref{G_uu}, 
we estimate $\langle T_{uu} \rangle$ at $r\gg a$ in the one-shell model (see Fig.~\ref{fig:Vaidya}).

First we investigate the behavior of a massless scalar field at $r\gg a$ in the Schwarzschild metric: 
\begin{equation}
ds^2 = - f(r)dt^2+\frac{1}{f(r)}dr^2+r^2d\Omega ^2 ,
\label{spherical metric}
\end{equation}
where $f(r)=1-\frac{a}{r}$. 
The action of the field $\varphi$ on this metric is 
\begin{eqnarray}
S&=&- \int d^4x \sqrt{-g}\left(\frac{1}{2} g^{\mu \nu} \partial _{\mu}\varphi \partial _{\nu}\varphi  \right) \nonumber \\
&=& - \frac{1}{2} \sum_{l,m}\int dt dr_{\ast} \varphi_{(l,m)} \left( \partial_t^2-\partial_{r_{\ast}}^2+V_l(r) \right)\varphi_{(l,m)} ,
\label{action of phi}
\end{eqnarray} 
where we have decomposed the field into partial waves 
\begin{equation}
\varphi (t,r,\Omega)=\sum_{l,m} \frac{\varphi_{(l,m)}(t,r)}{r} Y_{l,m}(\Omega) , \nonumber
\end{equation}
and introduced the new coordinate $dr_{\ast}\equiv \frac{dr}{f}$ and the effective potential for each partial wave with angular momentum $l$ as 
\begin{equation}
V_l(r)=f(r)\left( \frac{l(l+1)}{r^2}+\frac{\partial _r f(r)}{r}\right)\sim \frac{l(l+1)}{r^2}~~\text{at}~r\gg a.
\label{potential}
\end{equation}
This implies that only the s-wave survives at $r\gg a$ 
because partial waves with $l>0$ have to tunnel their own centrifugal barrier with the rate $P_l\sim e^{-l}$. 

Then let's consider the wave equation for scalar field $\varphi$ on the Vaidya metric \eqref{Vaidya}:
\begin{eqnarray}
 0&=& \nabla ^2 \varphi(x) \nonumber \\
 &=& \left( -2 \partial_u + \partial_r \frac{r-a(u)}{r} \right)\partial_r \varphi + 
     \frac{2}{r}\left( - \partial_u +  \frac{r-a(u)}{r} \partial_r \right)\varphi 
      - \frac{\hat l ^2}{r^2}\varphi ,
\end{eqnarray}
where $\hat l ^2$ is the Laplacian for angular directions. 
Here we take only the s-wave  
\begin{equation}
\varphi (u, r, \Omega)\approx \frac{e^{i\frac{\psi(u,r)}{\hbar}}}{r},
\end{equation}
and use the Eikonal approximation $(\hbar \rightarrow 0)$. 
Then we get 
\begin{equation}
0=\left( \frac{r-a}{r}\partial_r\psi - 2 \partial_u \psi \right)\partial_r \psi~.
\end{equation}
Therefore, for the outgoing modes, we obtain the equation: 
\begin{equation}
\partial_r \psi =0.
\label{sol}
\end{equation}

\begin{figure}
 \begin{center}
 \includegraphics*[scale=0.2]{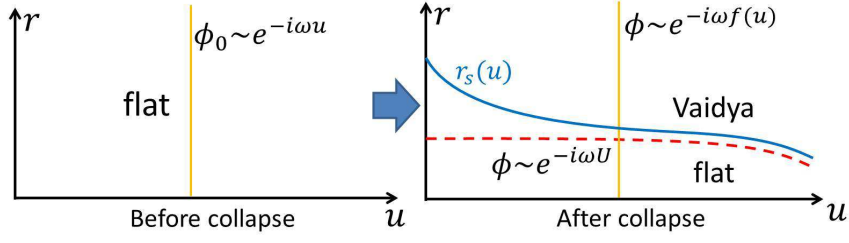}
 \caption{The time evolution of the Heisenberg operator $\phi$ before and after the collapse in the Eikonal approximation.}
 \label{fig:flux}
 \end{center}
 \end{figure}

Next in this approximation we consider time evolution of the Heisenberg operator $\phi$ at $r\gg a$ in the collapsing spacetime.
(See Fig.~\ref{fig:flux}.)
Before the collapse, the field is given by the spherical waves:
\begin{equation}
\phi_0(u,r)=\int_0^{\infty}\frac{d\omega}{2\pi}
\left( \frac{e^{-i\omega u}}{\sqrt{4\pi \omega} r}a_{\omega} + \frac{e^{+i\omega u}}{\sqrt{4\pi \omega} r}a^{\dagger}_{\omega}  \right),
\label{before}
\end{equation}
which corresponds to the field on the flat space.
Here the vacuum is defined as the Minkowski vacuum: 
\begin{equation}
a_{\omega}|0\rangle =0 \; \text{for} \; \omega>0.
\label{vac}
\end{equation}
After the collapse, the field becomes, from \eqref{sol},
\begin{equation}
\phi(u,r)=\int_0^{\infty}\frac{d\omega}{2\pi}
\left( \frac{e^{-i\omega f(u)}}{\sqrt{4\pi \omega} r}a_{\omega} + \frac{e^{+i\omega f(u)}}{\sqrt{4\pi \omega} r}a^{\dagger}_{\omega}  \right),
\label{after}
\end{equation}
where $f(u)$ is any increasing function of $u$.
We are here using the Eikonal approximation, so the phase remains constant on the outgoing mode:
\begin{equation}
f(u)=U(u)=-2r_s(u),
\label{phase}
\end{equation}
where $U$ is the time coordinate in the flat space inside the shell, 
and at the second equality we have used the junction condition on the locus of the shell \eqref{junction}.

Let's estimate the flux based on the above analysis.
We use the point-splitting regularization technique to subtract the divergence \cite{Birrell}:
\begin{equation}
\langle 0|:T_{uu}(u):|0\rangle=\lim_{u'\rightarrow u}\left[\langle 0|:\partial_u \phi(u) \partial_u \phi(u'):|0\rangle-
\langle 0|:\partial_u \phi_0(u) \partial_u \phi_0(u'):|0\rangle\right],
\label{T_uu1}
\end{equation}
where the time $u$ is after the collapse, and the r-dependence is not explicitly written because $r\gg a$.
First we introduce 
\begin{equation}
u'=u+\epsilon, \; u''=u-\epsilon~,
\end{equation}
and expand the equation with respect to $\epsilon$. 
By using \eqref{after}, the first term in \eqref{T_uu1} is estimated as 
\begin{eqnarray*}
&&4\pi r^2 \langle 0|:\partial_u \phi(u') \partial_u \phi(u''):|0\rangle\\ 
&=&\hbar \int^{\infty}_0 \frac{d\omega}{2\pi}\omega \dot f(u+\epsilon)\dot f(u-\epsilon)e^{-i\omega[f(u+\epsilon)-f(u-\epsilon)]}\\
 &=& \hbar \int^{\infty}_0 \frac{d\omega}{2\pi}\omega [\dot f+\ddot f\epsilon+\dddot f \epsilon^2/2+\cdots ]
 [\dot f-\ddot f\epsilon+\dddot f \epsilon^2/2+\cdots]e^{-i\omega[2\dot f\epsilon+\dddot f\epsilon^3/3+\cdots ]}\\
 &=& \hbar \int^{\infty}_0 \frac{d\omega}{2\pi}\omega [\dot f^2-\epsilon^2(\ddot f^2-\dot f\dddot f)+{\cal O}(\epsilon^4) ]
 e^{-2i\omega \dot f\epsilon}[1-i\omega \dddot f\epsilon^3/3+{\cal O}(\epsilon^5) ]\\
 &=&-\frac{\hbar }{8\pi \epsilon^2}+\frac{\hbar }{8\pi}\left[ \frac{\ddot f^2}{\dot f^2}-\frac{2\dddot f}{3\dot f}\right]+ {\cal O}(\epsilon^2),
\end{eqnarray*}
where $f=f(u)$. 
In the same way, the second term in \eqref{T_uu1} is estimated as 
\begin{equation*}
4\pi r^2 \langle 0|:\partial_u \phi_0(u') \partial_u \phi_0(u''):|0\rangle
=-\frac{\hbar }{8\pi \epsilon^2}~.
\end{equation*}
Thus, by using \eqref{phase}, we obtain the flux formula for $J(u)=4\pi r^2\langle 0|:T_{uu}(u):|0\rangle$ as 
\begin{equation}
J(u)=\frac{\hbar}{8\pi}\left[ \frac{\ddot U(u)^2}{\dot U(u)^2}-\frac{2\dddot U(u)}{3\dot U(u)}\right]\equiv \frac{\hbar}{8\pi}\left\{u,U\right\},
\end{equation}
which is \eqref{J}.

\section{Derivation of the Planck distribution without horizon}\label{sec:Planck}
We emphasize that the Planck distribution can be obtained even if the geometry has no horizon. 
All that is necessary is that 
the affine parameters on the null generators of past and future null infinity are related exponentially \cite{Barcelo:2010xk}. \ 

In this appendix, we will show that 
in our model, the expectation value of the number of the particle creation takes 
the form of the Planck distribution with the Hawking temperature $T_H(u)=\frac{1}{4\pi a(u)}$, 
in which $a(u)$ changes so slowly that \eqref{a_slow} holds.

We start with reviewing the standard calculation of the Hawking radiation. 
We consider the state in the Heisenberg picture that is annihilated by the positive frequency operators in the past infinity $a_\omega$: 
\begin{equation}
 a_\omega |0\rangle = 0 , \ \omega>0 .
\end{equation}
Because the profile of the wave is modified by the gravitational potential, 
the positive frequency operators in the future infinity $b_\omega$ 
is a superposition of the positive and negative frequency operators in the past infinity
$a_{\omega'}$, 
\begin{equation}
 b_\omega 
  = \int_{-\infty}^{\infty} d\omega' 
  A_{\omega \omega'} a_{\omega'}
  = \int_{0}^{\infty} d\omega' 
  \left(
   A_{\omega \omega'} a_{\omega'}
   + A_{\omega,- \omega'} a_{\omega'}^\dagger
  \right) . 
\end{equation}
Then in the future infinity the number operator takes the non-trivial value
\begin{equation}
 \langle 0| b_{\omega}^\dagger b_{\omega} | 0 \rangle
  = \int^{\infty}_0 d\omega'\,\left|A_{\omega,-\omega'}\right|^2 . 
  \label{NumOp}
\end{equation}
The coefficient $A_{\omega,-\omega'}$ is given by the Klein-Gordon inner product: 
\begin{equation}
A_{\omega,-\omega'}=(\varphi_{b}(u;\omega),\varphi_{a}^*(U;\omega')), 
\label{A}
\end{equation}
where $\varphi_{b}(u;\omega)$ is the wave function of the outgoing mode on the future null infinity and 
$\varphi_{a}(U;\omega)$ is that on the past null infinity. 

As in \ref{sec:der_flux}, we will use the Eikonal approximation for the s-wave.
Then, from \eqref{before} and \eqref{after}, 
\begin{equation}
\varphi_{b}(u;\omega) \sim \frac{1}{\sqrt{4\pi \omega}}\frac{e^{-i\omega u}}{r},~~\varphi_{a}(U;\omega) \sim \frac{1}{\sqrt{4\pi \omega}}\frac{e^{-i\omega U}}{r}.
\end{equation}
Then \eqref{A} becomes 
\begin{equation}
A_{\omega,-\omega'} = \frac{1}{2\pi} \sqrt{\frac{\omega}{\omega'}} \int^{\infty}_{-\infty}du e^{-i\omega u}e^{-i\omega' U(u)} .
\label{A2}
\end{equation}
Here we need the relation between $u$ and $U$.

(1) In the case of a single shell, 
we can use $U(u)=-2r_s(u)$ and $r_s(u)\approx a(u_*) + Ca(u_*)e^{-\frac{u-u_*}{2a(u_*)}}$ 
where $u_*$ is a time when the exponential factor remains. 
Here note that $C$ is positive because $r_s>a$.
Then we obtain 
\begin{equation}
U(u)=-2a(u_*)-2Ca(u_*)e^{-\frac{u-u_*}{2a(u_*)}}.
\end{equation}
Thus \eqref{A2} can be evaluated as 
\begin{equation}
A_{\omega,-\omega';u_*} = \frac{1}{2\pi} \sqrt{\frac{\omega}{\omega'}} \int^{\infty}_{-\infty}du e^{-i\omega u}e^{2i\omega' Ca(u_*)e^{-\frac{u-u_*}{2a(u_*)}}} ,
\label{A3}
\end{equation}
where the irrelevant phase factor is dropped.
Here the contribution from $u$ away from $u_{\ast}$ is negligible \cite{Hawking:1974sw, Barcelo:2010xk} 
because the only interval $[u_*-ka_*,u_*+ka_*]$ contributes to the integral, where $k$ is a constant $\sim 1$. 
After performing the $u$-integration, 
we obtain 
\begin{equation}
A_{\omega,-\omega';u_{\ast}} = \frac{a(u_*)}{\pi} \sqrt{\frac{\omega}{\omega'}} 
e^{-\pi a(u_*)\omega}\Gamma (i2a(u_*)\omega), 
\label{A4}
\end{equation}
where the irrelevant phase factor is omitted. 

(2) In the case of the asymptotic region \eqref{r_asy} of the continuous matter, 
we first expand $a(u)$ around $u_*$ which is a time in the region:
\begin{equation}
a(u)=a(u_*)+\dot a(u_*)(u-u_*)=a(u_*)-\frac{(u-u_*)}{a(u_*)^2},
\end{equation}
where for simplicity we have normalized the Hawking radiation as $\dot a =-\frac{1}{a^2}$. 
Then the redshift factor is estimated as 
\begin{equation}
\frac{dU}{du}=e^{-\int^{a(u)}_0\frac{da'}{\rho(a')}}=e^{-\frac{1}{4}a(u)^2}=e^{-\frac{1}{4}a(u_*)^2}e^{+\frac{u-u_*}{2a(u_*)}}.
\end{equation}
Thus we obtain
\begin{equation}
U(u)=\text{const.}+2a(u_*)e^{-\frac{1}{4}a(u_*)^2}e^{+\frac{u-u_*}{2a(u_*)}}\equiv D_* + 2C'a(u_*)e^{+\frac{u-u_*}{2a(u_*)}},
\end{equation}
where $C'>0$, and have 
\begin{equation}
A_{\omega,-\omega';u_*} = \frac{1}{2\pi} \sqrt{\frac{\omega}{\omega'}} \int^{\infty}_{-\infty}du e^{-i\omega u}e^{-2i\omega' C'a(u_*)e^{\frac{u-u_*}{2a(u_*)}}}
= \frac{1}{2\pi} \sqrt{\frac{\omega}{\omega'}} \int^{\infty}_{-\infty}du e^{i\omega u}e^{-2i\omega' C'a(u_*)e^{-\frac{u-u_*}{2a(u_*)}}}.
\label{A5}
\end{equation}
This is different from \eqref{A3} in the sign of the exponentials, 
but this integral leads to almost the same result:
\begin{equation}
A_{\omega,-\omega';u_{\ast}} = \frac{a(u_*)}{\pi} \sqrt{\frac{\omega}{\omega'}} 
e^{-\pi a(u_*)\omega}\Gamma (-i2a(u_*)\omega). 
\end{equation}

For the both cases, using the formula 
 \begin{equation}
\left | \Gamma(ix) \right|^2=\frac{\pi}{x\sinh (\pi x)}, 
\end{equation}
we obtain 
\begin{equation}
\left|A_{\omega,-\omega';u_{\ast}}\right|^2 = \frac{a(u_*)}{\pi \omega'} \frac{1}{e^{\frac{\omega}{T(u_{\ast})}}-1}, 
\label{A6}
\end{equation}
where 
\begin{equation}
T(u_{\ast})=\frac{1}{4\pi a(u_{\ast})} .
\end{equation}
By considering a wave packet around $u_{\ast}$, 
we arrive at the Planck-distributed Hawking radiation 
with temperature $T(u_{\ast})$: 
\begin{equation}
\langle 0| b_{\omega}^\dagger b_{\omega} | 0 \rangle _{u_{\ast}}=  \frac{1}{e^{\frac{\omega}{T(u_{\ast})}}-1}. 
\label{Planck}
\end{equation}

\section{Derivation of the stationary metric \eqref{metric_t}}\label{sec:metric}
We will derive the stationary metric \eqref{metric_t}.
The metric \eqref{Vaidya'} represents a vicinity around a point $(u',r')$ just outside a shell.
$r'$ is so close to $a'$ that, from \eqref{rho''}, 
\begin{equation}
\rho'=r'-a'=\frac{Nl_p^2}{24\pi a'}\approx \frac{Nl_p^2}{24\pi r'}~. 
\end{equation}
From \eqref{J'}, \eqref{rho2} and \eqref{J'3}, $\rho'{}^{-1}$ is also estimated as 
\begin{equation}
\frac{1}{\rho(a')}\approx \frac{24\pi r'}{N l_p^2}+\frac{2}{r'}~.
\end{equation}
By using these, the time coordinate $u'$ is related to that around the origin $U$ as 
\begin{equation}
\frac{du'}{dU}=\exp \left(\int^{a'}_{\sqrt{N}l_p}\frac{d\bar a}{\rho(\bar a)}\right) \approx \frac{r^{'2}}{Nl_p^2}e^{\frac{12\pi}{Nl_p^2}r^{'2}}~.
\end{equation}

From these, \eqref{Vaidya'} can be rewritten as 
\begin{eqnarray}
 ds^2 &=& -\frac{\rho}{r}\left(\frac{du'}{dU}\right)^2 dU^2 - 2 \frac{du'}{dU}dUdr+r^2d\Omega^2\nonumber \\
 &\approx& -\frac{r^2}{24\pi Nl_p^2}e^{\frac{24\pi}{Nl_p^2}r^2}dU^2 -2 \frac{r^2}{Nl_p^2}e^{\frac{12\pi}{Nl_p^2}r^2}dUdr + r^2d\Omega^2,
\end{eqnarray}
where we have replaced $r'$ with $r$.
Here we introduce the time coordinate around the origin:
\begin{equation}
dT=dU+24\pi e^{-\frac{12\pi}{Nl_p^2}r^2}dr,
\end{equation}
and express the metric as 
\begin{equation}
ds^2=-\frac{r^2}{24\pi Nl_p^2}e^{\frac{24\pi}{Nl_p^2}r^2}dT^2 + \frac{24\pi r^2}{Nl_p^2}dr^2 + r^2 d\Omega^2.
\label{metric_T}
\end{equation}

Now we connect it to the outside metric, that is, the Schwarzschild metric 
at the outermost shell $r=a+\frac{Nl_p^2}{24\pi a}$:
\begin{eqnarray}
ds^2&=&-\left(1-\frac{a}{r}\right)dt^2+ \left(1-\frac{a}{r}\right)^{-1}dr^2 + r^2d\Omega^2 \nonumber \\
 &=& -\frac{\rho}{r}dt^2+ \frac{r}{\rho}dr^2 + r^2d\Omega^2\nonumber \\
 &\approx&  -\frac{Nl_p^2}{24\pi a^2}dt^2+ \frac{24\pi a^2}{Nl_p^2}dr^2 + r^2d\Omega^2.
 \label{Sch}
\end{eqnarray}
Comparing this with \eqref{metric_T} at $r=a+\frac{Nl_p^2}{24\pi a}\approx a$, 
we obtain the relation
\begin{equation}
dT=\frac{Nl_p^2}{a^2}e^{-\frac{12\pi}{Nl_p^2}a^2}dt.
\end{equation}
Therefore we reach the metric
\begin{equation}
ds^2=-\frac{Nl_p^2r^2}{24\pi a^4}e^{-\frac{24\pi}{Nl_p^2}(a^2-r^2)}dt^2 + \frac{24\pi r^2}{Nl_p^2}dr^2 + r^2 d\Omega^2.
\end{equation}


\end{document}